\documentclass[letter,pre,showpacs,preprintnumbers,floatfix,twocolumn]{revtex4}

\usepackage{color}
\usepackage{graphicx}
\usepackage{subfigure}
\usepackage[]{amssymb}
\usepackage[]{amsmath}
\usepackage{ulem}

\newcommand{\Oh}{Oh}
\newcommand{\Ohd}{Oh_d}
\newcommand{\Ohb}{Oh_b}
\newcommand{\We}{W\!e}

\newcommand{\Wei}{W\!e_i}

\newcommand{\hmax}{h}

\newcommand{\hreb}{h}

\newcommand{\Y}{Y}
\newcommand{\G}{\varGamma}
\newcommand{\Gb}{\varGamma_b}

\newcommand{\Gpdb}{\varGamma_{(2,2)}}

\begin{document}

\title{The role of the droplet deformations in the bouncing droplet dynamics}

\author{\textbf{D. Terwagne,\textit{$^{a}$} F. Ludewig,\textit{$^{b}$} N. Vandewalle,\textit{$^{b}$} and
S.Dorbolo\textit{$^{b}$}}}\vspace{0.5cm}

\affiliation{\textit{$^{a}$~Department of Civil and Environmental Engineering, Massachusetts Institute of Technology, 02139 Cambridge, USA}}
\affiliation{\textit{$^{b}$~GRASP, Department of Physics, University of Li\`ege, B-4000 Li\`ege, Belgium. }}

\begin{abstract}
Droplets bouncing on a vibrated liquid bath open ways to methods of manipulating droplets, creating double emulsion and performing  pilot wave model experiments. In this work, we focus on the role of the droplet deformations in the vertical bouncing dynamics by neglecting the deformation of the surface of the bath. To be under this favorable conditions, low viscous oil droplet are dropped over a highly viscous oil bath. These droplets bounce vertically on the surface of the bath and exhibit many periodic trajectories and resonant modes when tuning the forcing parameters, i.e. the oscillation of the bath. This complex dynamics emphasizes the interplay between elastic energy storage and energy dissipation in droplets at each bounce. We propose to model droplets using a bouncing mass-spring-damper system that mimics a deformable droplet bouncing on a non-deformable liquid bath. From the experimental measurements, we constructed bifurcation diagrams of the bouncing trajectories and challenged our bouncing spring model. The agreement between experiment and the spring model reveals that this model can be used to rationalize and predict a variety of bouncing droplets behaviors involving multi-periodicities.
\end{abstract}


\maketitle

\section{Introduction}

A droplet laid onto a vertically vibrated liquid bath can, under certain conditions, bounce periodically and indefinitely. Periodic bouncing modes are of crucial importance in a variety of bouncing droplets experiments. These droplets can present interactions at a distance \cite{eddi2011vibration,eddi2009archimedean,eddi2008ratchet,protiere2006association}, deform following resonant modes which can sometimes lead to an horizontal self-propulsion mode \cite{dorbolo:2008}. In fact, one of the most spectacular behavior is, under particular conditions of the vibration, the transformation of the vertical bouncing into an horizontal persistent motion of the droplet \cite{couder2005nature}. This motion is due to the interaction between the droplet and the waves generated at the bath surface. This intriguing behavior reminds of the wave-particle duality of the matter. Actually, up to now, the experiments designed by analogy to classical quantum experiments demonstrate a disconcerting similitude between the macro world of droplets and the micro world of electrons and atoms \cite{couder:2006,eddi2009tunnel,fort2010memory}. These really interesting behaviors rely on the periodicity of the bouncing droplet modes which can be multi-periodic or chaotic.

The bouncing dynamics was shown to be strongly dependent on the relative bath/droplet deformation. Recent works by Mol\'a{\v{c}}ek and Bush \cite{molavcek2013bouncing,molavcek2013walking} related a precise and detailed model for the bouncing droplet. Considering that a quasi-static deformation of the droplet during the ``contact'' time with the bath occurs and  that the bath and the droplet deform of the same amplitude, the Authors discovered that a bouncing droplet can be modeled by a logarithmic spring. This model allows to reproduce very nicely the experimental results and allows for the first time to propose a complete explanation for the particular state of the bouncing droplet, the walker. The particularity of their model is that both the droplet and the bath deformations are modeled by a single logarithmic spring. This approach contrasts with the previous ones that considered the modeling of the deformation of a single droplet by a linear spring \cite{okumura2003water}. Indeed, the linear spring seems only natural as the deformation of a droplet follows spherical harmonic. It can be easily demonstrated that the increase of surface energy due to a small deformation along the first mode $Y_0^2$ is proportional to the square of the deformation, i.e. the droplet should behave like an harmonic oscillator \cite{hub}. Consequently, this question is rather interesting: when should the linear or the logarithmic spring be used? 

In this study, we will determine the precise role of the droplet deformation in the bouncing mechanisms when the bath deformation does not play any significant role. More precisely, we simplified the system by inhibiting the bath deformation and studied the role of the droplet deformation on the emergence of complex modes. In order to obtain this particular conditions, we focus on small droplets bouncing on a highly viscous liquid bath that is vertically vibrated. This system is certainly less complex than an homogenous system for which the droplet and the bath are made of the same low viscous oil. Indeed, in our system, no walker can be observed. In this case, the effect of the droplet deformation on the bouncing modes can be isolated and experimentally observed, these modes will be represented using bifurcation diagrams. We show that a simple model like the bouncing ball is unable to explain the diversity of the observed bouncing modes. We thus propose a model with the minimum physical ingredients required to capture the main features observed experimentally. Indeed, we demonstrate that a bouncing droplet can be modeled as a bouncing Kelvin-Voigt material (mass-spring-damper system) which is able to reproduce the trajectories of the droplet center of mass and their numerous periodicities when bouncing on a vibrating bath. The linear spring model is able to rationalize and predict a variety of bouncing droplet modes on a non-deformable liquid bath that is oscillating vertically. This shall lead to an alternative path to future models that would incorporate the bath deformation by, for example, using a second linear spring.

We will start this paper by giving, in section \ref{sec:Background},  a global view of the role of deformations in the bouncing droplet problem. Then, in section \ref{sec:ExpSetup}, we will describe the experimental setup and present an experiment of a droplet bouncing on a vibrating bath for a chosen frequency. The analysis of the periodicity of the trajectories as a function of the control parameter, namely the maximum acceleration of the plate, will be detailed in section \ref{sec:ExpRes}. These results will be summarized using bifurcation diagrams. In section \ref{sec:Modeling}, we will propose a minimal model based on a mass-spring-damper system to reproduce our experimental observations. Numerical simulations are performed to test the validity of the model. The numerical results are then presented in bifurcation diagrams which are compared to the experimental ones. Conclusions will close this paper.

\section{Background} \label{sec:Background}

In a series of recent works, bouncing modes have shown to exert particular features in bouncing droplet experiments. In both the pioneering work of Couder \textit{et al.} \cite{couder2005bouncing} on the bouncing threshold and of Dorbolo \textit{et al.} \cite{dorbolo:2008} on the various deformation modes, droplets are observed to bounce in phase with the vibrated bath.  Couder \textit{et al.} \cite{couder2005nature} discovered a self-propelled droplet, called `walker', guided by the surface waves generated by the bounce on the surface of the bath. This peculiar behavior is made possible when droplets double there bouncing period. Under these conditions, the droplets are able to trigger Faraday surface waves and use them to propel themselves horizontally. By controlling the bouncing modes and consequently the surface waves produced by the bounce, Eddi \textit{et al.} were capable of making different geometrical lattices of bouncing droplets \cite{eddi2009archimedean} and making these bonded droplets move on the surface as ratchets \cite{eddi2008ratchet}.

\begin{figure*}[htbp]
\begin{center}
\includegraphics[width=16cm]{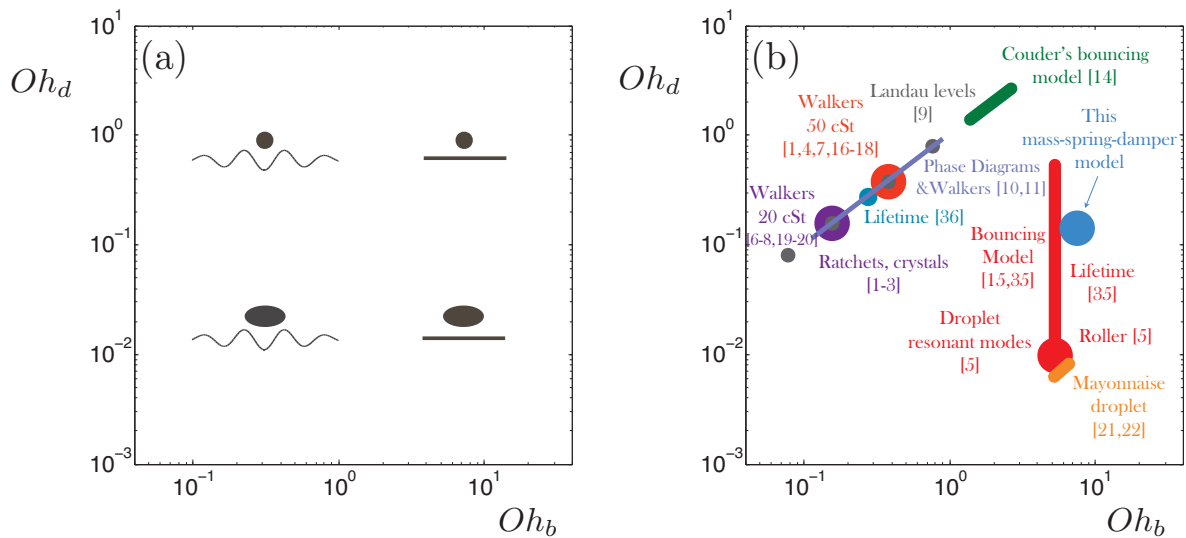}
\end{center}
\caption{(Color online) (a) The asymptotic behavior of a bouncing droplet on a vibrating liquid bath is categorized and sketched as a function of  the Ohnesorge parameters $\Oh_d$ of the droplet and $\Oh_b$ of the bath. These parameters represent the relative damping of the oscillations on the droplet and on the bath. (b) A selection of relevant works on bouncing droplets are plotted on the Ohnesorge diagram revealing the importance of the deformation of the droplet and/or of the bath in each experiment.\label{fig:DiagVisco}}
\end{figure*}

Let us strive to have a global view of the problem. A liquid droplet of diameter $D$ bounces on a liquid interface oscillating sinusoidally at a frequency $f$ and an amplitude $A$. At each bounce the droplet generates, on the liquid bath, a crater of characteristic length scale $D$, the diameter of the drop. The bath and the drop liquids are characterized by kinematic viscosities $\nu_{b}$ and $\nu_{d}$, densities $\rho_{b}$ and $\rho_{d}$, and surface tensions $\sigma_{b}$ and $\sigma_{d}$ , respectively. In earlier works \cite{dorbolo:2008, gilet:2008}, we suggested that a relevant parameter is the Ohnesorge number that compares viscous and capillary times. We then define Ohnesorge numbers of the bath $Oh_b=\nu_b \sqrt{\rho_b / (\sigma_b D)}$ and of the droplet $Oh_d=\nu_d \sqrt{\rho_d /( \sigma_d D)}$. These Ohnesorge numbers indicate how oscillations are damped by viscosity on the bath and on the droplet. Thus, these numbers give an indication on the relative importance of the bath and droplet deformations in experiment. On Fig. \ref{fig:DiagVisco}a, we have represented a sketch illustrating the importance of these deformations as a function of the Ohnesorge numbers $Oh_b$ and $Oh_d$.  Obviously, both parameters are not sufficient to describe univocally the behaviors of the droplets. This map on Fig.~\ref{fig:DiagVisco}a can be seen as a projection of the muti-dimension phase diagram on a 2D map defined by $\Ohb$ and $\Ohd$.

On Fig. \ref{fig:DiagVisco}b, we have situated a selection of bouncing droplet studies on the deformations map. The plotted areas were adjusted for a better visualization. The first model for the bouncing threshold has been developed by Couder \textit{et al.} \cite{couder2005bouncing} for highly viscous droplets on a highly viscous bath, both deformations of the bath and of the droplet are  rapidly damped (top right corner of the map). Using oil of viscosity 50~cSt, Proti\`ere \textit{et al.} investigated the interactions of droplets through the capillary waves they emitted and initiated fundamental studies on the walker's mechanism \cite{couder2005nature,protiere2005organization,protiere:2008,protiere2006orbital,protiere2006association}.  Eddi \textit{et al.} performed pilot wave experiments using walkers made of 20~cSt silicone oil \cite{eddi2009tunnel,eddi2011memory,eddi2012level}. These series of experiments are based on small droplets (walkers) emitting waves on the liquid surface, these works are thus located on the left upper part of the map (cf. Fig.~\ref{fig:DiagVisco}b). Experiments on ratchet motion, drifting rafts, Archimedean lattices and oscillations in crystal of droplets were performed with the same set of experimental conditions \cite{eddi2011vibration,eddi2009archimedean,eddi2008ratchet}. Fort \textit{et al.} changed the viscosity of the droplet and the bath in order to change the length of the Faraday waves emitted by the walkers \cite{fort2010memory}, these experiments on the Landau levels are located in grey around the region described previously. Finally, the systems investigated by Bush \textit{et al.} \cite{molavcek2013bouncing, molavcek2013walking} are located around the walkers' region. Dorbolo \textit{et al.} \cite{dorbolo:2008} focused more on the droplets deformation and used a highly viscous liquid bath to inhibit any deformation of the bath. We showed that a resonance phenomenon occurs between the bouncing droplets mode of deformation and the frequency of the vibrated bath, droplets deform as spherical harmonics which modes are selected by the forcing parameters. These deformation modes have a significant influence on the acceleration threshold for bouncing and this has been rationalized by an analytical model  \cite{gilet:2008}. We also evidenced that deformation of the droplet can be used to make droplets move horizontally on the surface of the bath  \cite{dorbolo:2008}. A non axisymmetric mode of deformation can be excited for specific forcing parameters which lead to self propulsion of the droplet on the bath. As well, these deformations can be used to create double emulsion in bouncing compound droplets~\cite{terwagne2010langmuir, terwagne2010chaos}. All these works relying specifically on the droplet deformation are situated in the lower right part of the map.

It is of importance to position our work in this map. As we want to study the role of the droplet deformation on the bouncing dynamics neglecting the bath deformation, we positioned our work at the level of the $\Ohd$ corresponding to the 20~cSt walkers but at a relatively high $\Ohb$ corresponding to a highly viscous bath (see disk  annotated `\textit{This mass-spring-damper model}' on Fig. \ref{fig:DiagVisco}b).

\section{Experimental setup} \label{sec:ExpSetup}

\subsection{Droplet generator}

We built a droplet dispenser to create submillimetric droplets of silicone oil (Dow Corning 200) of kinematic viscosity 20~cSt. The silicone oil surface tension is $\sigma=20.6\,\rm{mN/m}$ and density is $\rho=949\,\rm{kg/m}^3$. On Fig. \ref{fig:DispExp}a, a small container, with a hole at the bottom, is closed by a piezoelectric chip at the top and  is filled with the silicone oil. By injecting a short electric impulse ($\sim 5$~ms) to the piezoelectric chip, a shock wave is released in the container and, through the hole, a drop is ejected with a diameter related to the size of the hole.  In this case, the hole as a diameter of 600~$\mu$m and we are able to produce droplets with diameters ranging from 650~$\mu$m to 750~$\mu$m depending on the intensity of the impulse sent to the piezoelectric chip. The droplet size was optically measured using a high speed camera (IDT-N3) and a macro lens. For a given intensity value of  the pulse, the dispersion of the diameters of the droplets that are generated stays within $3\%$. With this technique, we are able to produce repeatedly droplets of silicone oil (and even water) of the same diameter in a range of 100~$\mu$m to a few millimeters by changing the hole diameter. For more information, see Ref \cite{Terwagne2011thesis}.

In order to generate droplets of different diameters, one can also merge bouncing droplets on the bath. This can be made by pushing them against each other using the meniscus that is created by a stick dipped in the bath. For our study of the droplet trajectories, we chose to work with droplets of diameter 890$\,\mu$m and made of 20~cSt silicone oil which are typical droplets used for the 20~cSt walking droplets \cite{couder2005nature} and pilot wave experiments \cite{couder:2006,eddi2009tunnel,eddi2011memory,fort2010memory}. Using the dispenser, we made two identical droplets of 700$\,\mu$m on the vibrating bath. Then we force them to coalesce together and the resulting droplet has a diameter of $D=890\,\mu$m.
\subsection{Vibrating bath}

The submillimetric droplets, of silicone oil 20~cSt, generated by the droplet dispenser described above, are gently laid on a highly viscous liquid bath made of 1000~cSt silicone oil. This bath is vertically vibrated using an electromagnetic shaker GW-V55 following a sinusoidal motion of frequency $f$ and amplitude $A$. We characterize the oscillation of the bath by the dimensionless forcing amplitude $\G=4\pi^2 A f^2/g$, where $g$ is the acceleration of the gravity.

An accelerometer (PCB-Piezotronics, 352C65), which delivers a tension proportional to the acceleration, is glued on the vibrating plate. With the latter linked to an oscilloscope, we measure the maximal dimensionless acceleration of the plate $\G$. All the acceleration values in this manuscript are given with an error of 2 \%.

The frequency of the oscillating bath is arbitrarily chosen to be $f=50$~Hz which is in the range of the forcing frequency commonly used in bouncing droplets experiments \cite{terwagne2010langmuir,dorbolo:2008,gilet:2008, eddi2011vibration,eddi2009archimedean, fort2010memory} and low enough to have a good time resolution while tracking with the high speed camera.

\subsection{Droplet tracking}

\begin{figure}[h!]
\begin{center}
\includegraphics[width=0.38\textwidth]{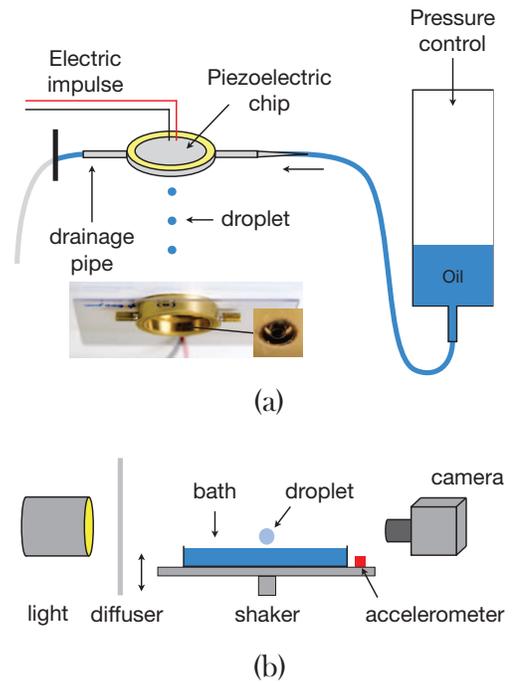}

\end{center}
\caption{(Color online) (a) Silicone oil droplets (20$\,\rm{cSt}$) are made using a droplet dispenser which consists in a small container with a hole at the bottom and a piezoelectric chip at the top. A short electric impulse is injected to the piezoelectric chip which produces a shock wave in the container that ejects a droplet through the hole at the bottom. (b) These droplets were laid on the surface of a highly viscous silicone oil bath (1000$\,\rm{cSt}$) that is vertically vibrated using an electromagnetic shaker. A high speed camera (1000 frames per second) recorded the motion and deformation of the drop from the side which were enhanced by a well positioned backlight. \label{fig:DispExp}}
\end{figure}

The bouncing droplets trajectories are tracked using the fast video camera N3 (1000 frames per second) from Integrated Device Technology, Inc. (IDT) (see Fig. \ref{fig:DispExp}b). As fast video cameras demand a powerful lighting,  we use several sets of 7LED-cluster from IDT. The position of the vibrating bath is tracked using a fixed wire glued to the plate which supported the bath. All the trackings, are made using the image analysis tracking tools provided by the software \textit{MotionProX} from IDT. Basically, the program allows us to precisely follow the center of mass of the droplet as it is bouncing. Note that as the bath is not deforming, droplets stay at all time entirely visible to the camera which simplifies the tracking. The recorded trajectories are then post-processed using \textit{Matlab}. As the droplet is never in contact with the bath and as the interaction time with the bath is of the order of 10 ms, a criterion had to be chosen to define when the droplet bounces. The minimum of the trajectory is defined as the rebound moment. Other choices as, for example, the inflexion points of the trajectory would have produced similar results as the presented ones. To have an estimation of the error on the measurements, the height of the bouncing droplet ranged from 0.5 to 3 mm and the error on the tracking method was estimated to be about 0.05~mm.

\section{Experimental results}\label{sec:ExpRes}

\subsection{Bouncing modes}

For a fixed frequency of the oscillating bath and a forcing acceleration $\G$ higher than a threshold $\Gb$, the droplet is observed to bounce periodically on the surface of the liquid bath as observed and described in previous publications \cite{couder2005bouncing, eddi2008ratchet, dorbolo:2008, gilet:2008}. At each bounce, the air film between the droplet and the surface of the bath is squeezed and can resist until the droplet bounces back into the air. This allows the air film to be renewed at each bounce. In some case, we can observe that droplets are permanently bouncing for acceleration lower than the gravity ($\Gb<1$). For our experimental set of parameters, droplets are simply bouncing at the threshold, i.e. one bounce per oscillation of the bath, and the droplet deformation can be described as a spherical harmonic $Y$ of degree 2 and mode 0 ($\Y^0_2$) \cite{dorbolo:2008, gilet:2008}. At each bounce the droplet looses energy in the droplet internal motion and in the lubrication film which are both eventually dissipated by viscosity. This loss at each bounce is compensated by the energy input coming from the oscillation of the bath. This energy compensation defines the threshold for bouncing. We should note that this mechanism defines also a bouncing threshold for air bubbles bouncing under an oscillating air/liquid interface, which experiment is exactly the negative of ours \cite{zawala2011bouncing}.

When $\G$ is further increased, above the acceleration threshold for bouncing $\Gb$, the droplets experience complex vertical trajectories that can be periodic over a few bath oscillations. Indeed, according to the oscillation phase, some extra energy can be used for a higher bounce and at the next bounce the energy given could be lower. However, on average the energy loss at each bounce has to be balanced after a number $p$ (an integer) of periods. The differences in bouncing heights are characterized by different flight times between successive bounces which can be measured. We investigated the bouncing droplet trajectories as a function of the forcing parameters $\G$ of the oscillation of the bath, the forcing frequency $f$ and the size of the drop $D$ being kept constant.

We investigate $890\,\mu m$ diameter drops of 20~cSt silicone oil bouncing on a 1000~cSt bath that oscillates vertically at $f=50$~Hz. The trajectories were recorded from the side with a fast video camera at 1000 fps. Stable bouncing modes are sought by varying the initial conditions of the bounce for bath accelerations between $\Gb = 0.9$ and $\G=5$ by steps of approximately 0.1. In each movie, we recorded about 50 bath oscillations. The center of mass of the droplet and the bath positions were tracked over time.

\begin{figure*}[htbp]
\begin{center}
\includegraphics[width=16cm]{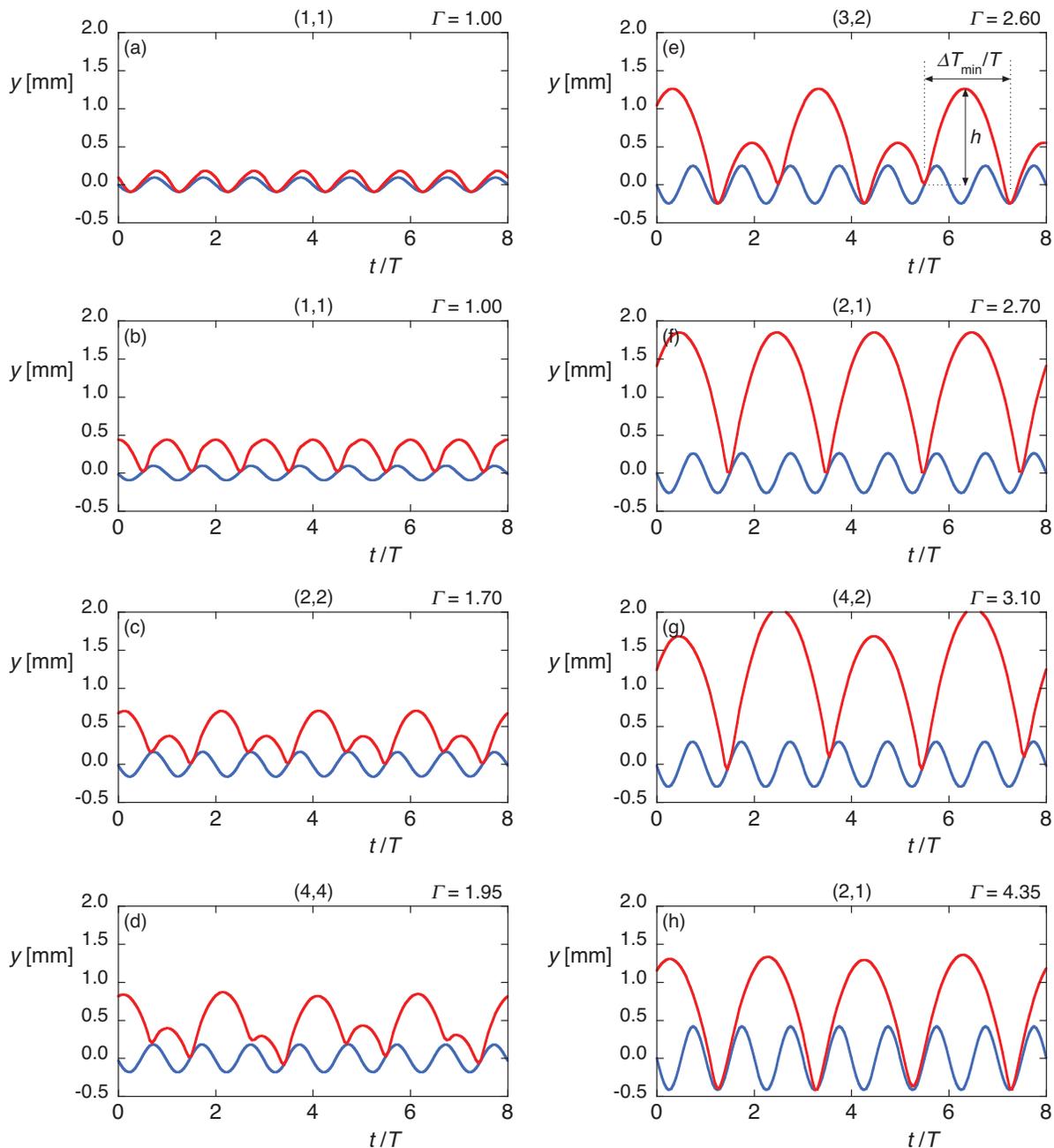}
\end{center}
\caption{\label{fig:Traj890um50Hz} (Color online) Experimental trajectories of a 20~cSt silicone oil droplet of diameter $D=890\,\mu$m bouncing on a bath oscillating at a frequency of 50~Hz for various accelerations $\G$. The bouncing mode $(p,q)$ and the forcing acceleration $\G$ are indicated on each figure. The time interval $\varDelta T_{\mathrm{min}}$ between two successive bounces and the bouncing heights $h$ that are measured on each trajectory are illustrated on graph (e).}
\end{figure*}

On Fig.~\ref{fig:Traj890um50Hz}, we present different characteristic periodical trajectories that corresponds to different forcing accelerations $\G$. On each graph, the lower sinusoidal curve (blue color online) correspond to the position of the bath while the other upper curve (red color online) represent the center of mass of the droplet. Trajectories of the center of mass of the droplet are vertically translated for better visualization. Note that the droplet is not bouncing instantaneously on the bath surface, it deforms itself when interacting with the bath that lasts about one third of a bath oscillation period. We define the droplet ``contact'' with the bath as the minimum in the trajectory of the center of mass of the droplet.

Notable periodic bouncing trajectories can be observed on Fig.~\ref{fig:Traj890um50Hz} with patterns that extend over a few bath oscillations and are periodically repeated. To characterize the different modes,  we refer them by two parameters $(p,q)$ according to the classical notification: the drop bounces $q$ times while the bath oscillates $p$ times.

\begin{figure*}[htbp]
\begin{center}
\includegraphics[width=16cm]{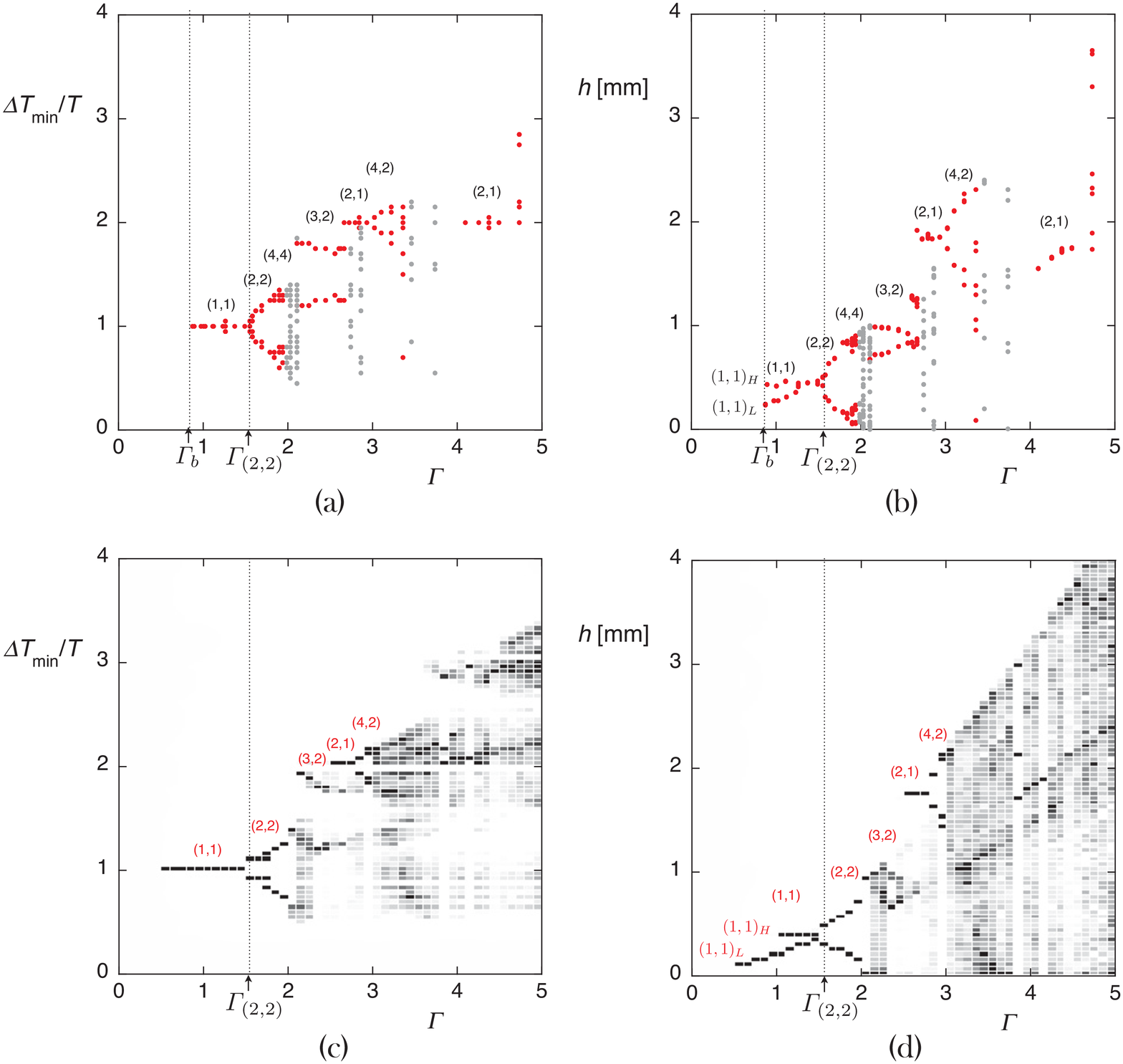}
\end{center}
\caption{\label{fig:DiagBifNumZETA50} (Color online) Trajectories characterization of the experiments and of the simulations ($k=0.072$~N/m $c=112~10^{-6}\,\rm{kg/s}$) for a droplet of diameter $890\,\mu$m bouncing on a rigid liquid bath oscillating at 50~Hz for various forcing accelerations $\G$. (a) Experimental measurements of the time intervals $\varDelta T_{\mathrm{min}}$, normalized by the oscillation period $T$ of the bath, between two successive bounces; and (b) experimental measurements of the bouncing heights (cf. Fig.~\ref{fig:Traj890um50Hz}e). (c) Simulation measurements of the time intervals $\varDelta T_{\mathrm{min}}/T$; and (d) simulation measurements of the bouncing heights $h$. The different bouncing modes $(p,q)$ are indicated on each diagram.}
\end{figure*}

As we will see, the bifurcation diagram gives more information on the bouncing modes than the ($p$,$q$) denomination. To construct the bifurcation diagram, we measured two parameters characteristics of each bounce: the flight time $\varDelta T_\mathrm{min}$ and the height of the bounce $\hreb$. We define $\varDelta T_\mathrm{min}$ as time intervals between two successive ``contacts'' of the droplet with the bath (that corresponds to two successive minima of the trajectory) and $\hreb$ as the height of successive bounces which is defined as the difference of height in between a maximum and the previous minimum of the droplet trajectory. Both parameters are illustrated on Fig.~\ref{fig:Traj890um50Hz}e. The measurements of $\varDelta T_\mathrm{min}$ are normalized by the oscillation period of the bath $T$.  We measured $\varDelta T_\mathrm{min}/T$ and $\hreb$ as a function of the forcing $\G$ on each of the experimental movies and reported these values as a function of $\G$ on Fig.~\ref{fig:DiagBifNumZETA50}a-b. On these figures, dark data points (red color online) are related to the droplet trajectories observed as periodic while the light grey data points are related to the chaotic trajectories, no periodic motion was observed on these movies. The bouncing modes $(p,q)$ are indicated above their corresponding  data points. These figures are called bifurcation diagrams as they depict the bouncing modes evolution as a function of the control parameter, $\G$.

\subsection{Bifurcation diagrams}

The bifurcation diagrams on Fig.~\ref{fig:DiagBifNumZETA50}a-b can be read as follow. A droplet permanently bounces on a bath oscillating at an acceleration $\G>\Gb$. At the bouncing acceleration threshold $\Gb$, the droplet bounces simply on the bath following a mode (1,1). When $\G$ is increased, the bouncing droplet reaches a bifurcation when $\G=\G_{(2,2)}$, above which two successive bounces become uneven, corresponding to the mode (2,2). After that, when $\G$ is increased further, a succession of bifurcations and modes (4,4), (3,2), (2,1) and (4,2) emerge and these modes can be interspersed with chaotic zones.

To facilitate our understanding of the bouncing modes, we analyze the bifurcation diagrams on Fig.~\ref{fig:DiagBifNumZETA50}a-b in parallel with the trajectories on Fig.~\ref{fig:Traj890um50Hz}. Indeed, these trajectories are represented on the diagram by one or  several data points depending on the complexity of the trajectories. The first bouncing mode (1,1) characterized by even bounces at each period of the oscillation of the bath appears to be stable in two different configurations. They differ in the phases and the heights of the bouncing (see Fig.~\ref{fig:Traj890um50Hz}a-b). We denote by (1,1)$_L$ the \textit{lower} bouncing trajectory that is in phase with the oscillation of the bath and with smaller bounces (see Fig.~\ref{fig:Traj890um50Hz}a). (1,1)$_H$ refers to the  \textit{higher} stable bouncing trajectory (1,1) that is out of phase with the bath oscillation and which bounces are higher (see Fig.~\ref{fig:Traj890um50Hz}b). As both trajectories are characterized by the same time interval between droplet ``contacts'' with the bath, they cannot be discerned in the time intervals phase diagram on Fig.~\ref{fig:DiagBifNumZETA50}a. However, they can be discerned on the bouncing height diagram on Fig.~\ref{fig:DiagBifNumZETA50}b.  The height of the (1,1)$_L$ type varies with $\G$ while the ``contact'' phase, observed on the trajectory plots, remains fixed. Conversely, the height of the (1,1)$_H$ type remains fixed while the ``contact'' phase adapts itself with increasing $\G$. This latter case, is much more similar to a bouncing ball behavior \cite{gilet:2009b}. As the acceleration increases, the droplet will continue to bounce in this (1,1) mode as long as the droplet can adjust the contact phase to compensate the loss of energy at each bounce. This locking of the phase is a well known feature in the bouncing ball phenomenon. However, the differentiation of the (1,1) mode into a high and a low mode is the consequence of the deformation of the droplet. This constitutes a main difference with the bouncing ball problem.

When $\G$ is increasing to $\G=1.70$ (see Fig.~\ref{fig:Traj890um50Hz}c), the droplet experiences a large bounce followed by a smaller one during two oscillations of the bath which mode is denoted by (2,2). When $\G$ is further increased, a wide variety of periodic modes can be observed. They are presented on Fig.~\ref{fig:Traj890um50Hz}d-h, the corresponding mode $(p,q)$ and forcing acceleration $\G$ are written at the top of each figure. These modes can be easily situated on the bifurcation diagrams on Fig.\ref{fig:DiagBifNumZETA50}a-b.

Another consequence of our deformable bouncing object is that the mode (2,1) is observed in two ranges of acceleration $\G \approx 3$ and $\G \approx 4.3$ (cf. Fig.~\ref{fig:DiagBifNumZETA50}a-b). Both modes described by the same parameters (2,1) are represented on Fig.~\ref{fig:Traj890um50Hz}f and Fig.~\ref{fig:Traj890um50Hz}h and occur for different $\G$.  They are characterized by a same time interval between the minima. However, the bouncing height of the mode occurring at higher $\G$ increases with $\G$ as long as the contact phase remains constant. The other mode (2,1) is characterized by a constant height for an increasing $\G$ when the contact phase varies.  This is qualitatively the same behaviors that is observed for the (1,1)$_H$ and (1,1)$_L$ modes. It is important to note that the mode (2,1) is known to be the characteristic mode of the walker droplets \cite{protiere2006association}. Even if walkers are observed on a low viscous (deformable) bath, the present work suggest that several (2,1) modes could exist for the walker.

From these observations, we can draw the following conclusions:
\begin{itemize}
\item For a same set of forcing parameters, multiple trajectories can be stabilized. On Fig.~\ref{fig:Traj890um50Hz}a and b, the mode  (1,1)$_H$ and (1,1)$_L$ can be observed for the same parameter $\G$, the selection of mode depending on the initial conditions. The main differences between both modes reside in the bouncing heights and the contact phases which are different.

\item The classical notification $(p,q)$ does not refer univocally to one trajectory. Indeed, as for the mode (1,1), the mode (2,1) describes two different kind of trajectories, the droplet bounces in phase and out of phase with the plate (see Fig.~\ref{fig:Traj890um50Hz}a-b for the mode (1,1) and Fig.~\ref{fig:Traj890um50Hz}f-h for the mode (2,1)). This shows that the phase of the rebound is an additional relevant parameter.

\item According to the initial conditions, a trajectory at a fixed $\G$ can be either stable or chaotic. As two different stable modes, e.g.  (1,1)$_H$ and (1,1)$_L$, can be selected depending on the initial condition, a chaotic or a stable mode can also be observed for a same set of parameters. Note that, under certain conditions, a droplet can alternatively pass from stable to chaotic and vice versa.
\end{itemize}

We thus observed that the bouncing trajectories for a droplet on a vibrated bath is complex and modes are numerous. The bouncing ball model is too simple and is not able to reproduce the variety of trajectories that are observed \cite{Tufillaro:1986, gilet:2009b}. This is mainly due to the fact that in the bouncing droplet case, the droplet can deform. This additional degree of freedom must be taken into account as it influences the impact time which is not instantaneous. In order to take this effect into account, we propose to come back to a more basic system to determine the inter-relation between the impact and the deformation. For this purpose, we will start by a detailed observation of a droplet bouncing on a static bath.

\section{Modeling}\label{sec:Modeling}

\subsection{Static liquid bath \label{subsec:ExoObs}}

\begin{figure}[htbp]
\begin{center}
\includegraphics[width=8cm]{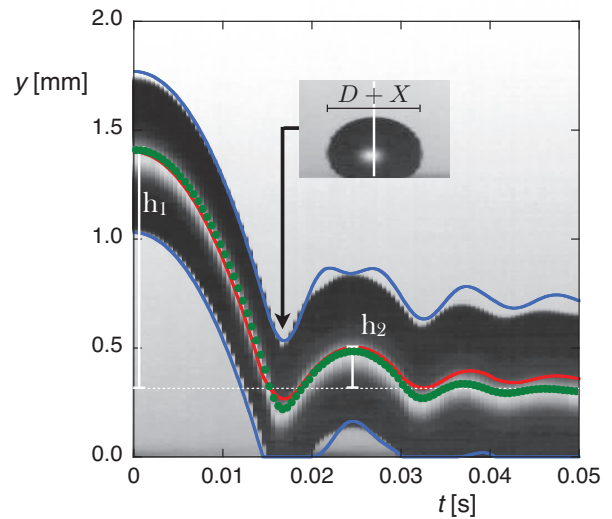}
\end{center}
\caption{\label{fig:DropBounce} (Color online) Spatio-temporal diagram of a 20~cSt silicone oil droplet of diameter $D=740\,\mu$m falling on a static highly viscous 1000~cSt silicone oil bath. Time elapses from left to right. The droplet experiences several bounces of heights $h_1$, $h_2$ that are measured from the center of mass of the droplet when it is floating (white dotted line). The impact speeds of the first bounce is about $0.15$~m/s ($\We \approx 0.8$). A snapshot illustrate the shape of the droplet when its deformation $D+X$ is maximal during the bounce. Successive dots (green color online) represent the center of mass of the droplet detected on the successive snapshots. The simulated trajectory of a falling mass-spring-damper system, from a height of $h_1=1.1$~mm, with parameters $k=0.072$~N/m and $c=112~10^{-6}\,\rm{kg/s}$, is superposed using 3 solid lines on the spatio-temporal diagram. The central solid line (red color online) is the trajectory of the center of mass of the system and both outer, upper and lower, solid lines (blue color online) are the trajectories of the point masses $m_1$ and $m_2$, respectively.}
\end{figure}

To gain insight on the bouncing mechanism of a droplet, we laid a droplet on the bath when it was at rest. We observed that the droplet bounces several times on the bath before coalescing with it. Using a high speed camera, we recorded an experiment, from the side, at 2000~fps (frames per second). In this case a 740 $\mu$m droplet was considered. The reason of this particular size is that we study the behavior of a single droplet made by the dispenser. On Fig.~\ref{fig:DropBounce}, a succession of the droplet bounces is detailed. Based on the sequence of images, we constructed a spatio-temporal diagram according to the following method: the drop vertical centerlines of each image are juxtaposed. The time elapses from left to right and the centre of mass of the droplet is measured and highlighted by successive dots (green color online).  On the figure, one snapshot illustrates the droplet when it is squeezed at maximum during the impact. In this snapshot, the white vertical line indicates the pixel line represented on the spatio-temporal diagram. Before impact, the drop is spherical and has a diameter $D$. Then, it deforms due to the impact. The maximal droplet horizontal extension is denoted by $D+X$. When deformations are small ($\We<1$) the droplet shape can be modeled by an oblate spheroid of major axis $D+X$ \cite{okumura2003water,richard2000bouncing} which surface increase is proportional to $X^2$. Then, the drop returns to a spherical shape and takes off, namely its center of mass goes up, without entering in contact with the underlying liquid. Residual oscillations on the drop surface are then rapidly damped by viscosity.

Physically, the droplet has an initial kinetic energy that is converted in surface energy during the bounce. The surface energy is then restored and converted into kinetic energy of the center of mass of the droplet and into kinetic energy of the internal liquid motion. The latter is eventually dissipated by the viscosity while the kinetic energy of the center of mass is converted into gravitational potential energy which is maximal when the maximal height of rebound $\hmax$ is reached.

Based on these observations, a model can be imagined to reproduce these bouncing modes. They are observed to be linked to the frequency and size of the droplets. A natural bouncing model would be the inelastic or semi-elastic bouncing ball which reproduces this kind of phase diagram. This model is characterized by an instantaneous contact solely governed by the restitution coefficient $\epsilon$ of the ball. In our case, due to the finite contact time of the bouncing droplet, this model is too simple to capture the complex interaction in between the droplet and the oscillating bath. Let us remember that the term ``contact'' is improper because the droplet and the bath are never in contact. A droplet is considered to take off when the lubrication film is sufficiently thick for the force it exerts to be neglected. Gilet and Bush \cite{gilet:2009a,gilet2009soap} faced the same issue as they observed droplets bouncing on a vibrating soap film. They developed a theoretical model based on the second Newton law to predict trajectories. Depending on the shape of the soap film, they could deduce the force acting on the droplet. Similarly, Eichwald \textit{et al.} \cite{eichwald:2010} investigated the phase diagrams of a solid bouncing ball on an elastic horizontal membrane whose frame oscillates. Note that in both works the bouncing motion is due to the deformation of the oscillating membrane or of the soap film and not from the deformation of the bouncing object.

In a previous paper \cite{dorbolo:2008}, we demonstrated that a droplet behaves as a forced harmonic oscillator, the surface tension providing the restoring force and the droplet viscosity the damping. For Weber number at impact $\Wei<1$, the droplet stores energy in surface energy which scales as $\sigma X^2$ \cite{okumura2003water}, where $X$ is the length increment of the droplet at impact (cf. Fig.~\ref{fig:DropBounce}). We can thus, naturally, propose a 1D model for the droplet as a system of two masses linked by a spring which is characterized by a stiffness $k$ in parallel to a dashpot with a damping coefficient $c$ (cf. Fig.~\ref{fig:spring}). The behavior of the spring system is governed by the Newton laws written for both masses. A  one spring model without a dashpot has already been applied to infer a bouncing criteria and a restitution coefficient to inviscid droplets impacting an hydrophobic surface at high speed (corresponding to high deformation) \cite{biance2006elasticity}. On our side, droplets bounce on a vibrating bath with Weber number $\Wei<1$ or close to 1. Thus, we will study this system when the surface energy that is of the same order of magnitude as the droplet kinetic energy.

In the next section, we will investigate the dynamics of the mass-spring-damper system bouncing on a static as well as on an oscillating plate. We started by numerically studying the behavior of a spring falling on a static plate. Then, we will compare the results to the experimental data (cf. Fig. \ref{fig:DropBounce}) in order to adjust the spring parameters $k$ and $c$. Finally, we will study the bouncing modes of this mass-spring-damper system bouncing on an oscillating plate. We will then make a comparison with the experimental bifurcation diagrams (cf. Fig.~\ref{fig:DiagBifNumZETA50}a-b). We will show that, this model captures the main features that are experimentally observed which cannot be retrieved using the bouncing ball system solely governed by the coefficient of restitution. Indeed, in the case of deformable object, the coefficient of restitution is not constant and depends strongly on the dynamics of the bounce.

\subsection{Bouncing spring model}

\begin{figure}[htbp]
\begin{center}
\includegraphics[width=6cm]{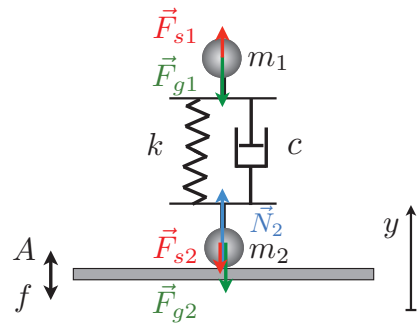}
\end{center}
\caption{\label{fig:spring} (Color online) The bouncing droplet is modeled by two masses $m_{1,2}$ linked by a spring of stiffness $k$ in parallel to a dashpot with a damping coefficient $c$. The spring system is in contact with the plate oscillating sinuosidally at a frequency $f$ and amplitude $A$. As $m_2$ is in contact with the plate, a normal force $\vec N_2$ acts from the plate on $m_2$. In this case the spring is compressed  ($\varDelta y<L$), both masses feel outwards spring forces $\vec F_{s1}$ and $\vec F_{s2}$ in addition to the gravity forces $\vec F_{g1}$ and $\vec F_{g2}$.}
\end{figure}

To reproduce the complex dynamics of the bouncing droplet, we model it by a Kelvin-Voigt material (mass-spring-damper system) bouncing on a oscillating rigid plate, as sketched on Fig.~\ref{fig:spring}. The coordinate $y$ is taken as the vertical one (positive upward), $y_1$ and $y_2$ are the coordinates of the upper mass $m_1$ and of the lower mass $m_2$. The spring is characterized by a stiffness $k$ and the dashpot by a damping coefficient $c$. At rest, the static length of the spring is $L$.

On Fig.~\ref{fig:spring}, we also represented the gravity forces $F_{g1}=-m_{1}g$ and $F_{g2}=-m_{2}g$ that act on both masses. The spring exerts also forces on each mass, they are $F_{s1}=-k(\varDelta y-L)$ and $F_{s2}=k(\varDelta y-L)$ with $\varDelta y=y_1-y_2$. We add a viscous dissipation force which is proportional to the speed of elongation of the spring, with a viscous damping coefficient of proportionality $c$. When the lower mass $m_2$ is in contact with the plate, it experiences a normal force $N_2$ from the plate.

The normal force $N_2$ depends in a complex manner on the elongation of the spring and on the bath acceleration which both vary during contact. Therefore we use molecular dynamics to compute this force $N_2$ when there is contact~\cite{cundall1979discrete,Opso2012}. This normal force is null when there is no contact.   Molecular dynamics simulation models numerically the contact between the plate and the mass $m_2$ as a damped spring interaction when it is in compression only. To model the fact that the plate is rigid, the mass of the plate $m_p$ is considered as a very large mass~($m_p \gg m_2$) and the stiffness of the contact spring $k_{c}$ is chosen high enough to avoid any interaction with the spring that models the droplet ($k_{c} \gg k$). Also, the damping coefficient modeling the contact is taken high enough to have an overdamped system modeling the contact. The value of these parameters, $m_p$ and $k_c$, does not influence the results as long as they respect the above inequalities~\cite{Taberlet2005}. The contact dynamics is thus modeled as a very stiff over damped spring which gives a good approximation of the normal force exerted by the rigid plate on the mass in contact.

Considering the droplet model, Newton laws can be written for both masses $m_1$ and $m_2$ as
\begin{align}
m_1 \ddot y_1&=-m_{1}g-k(y_1-y_2-L)-c(\dot y_1-\dot y_2), \\
m_2 \ddot y_2&=-m_{2}g+k(y_1-y_2-L)+c(\dot y_1-\dot y_2)+N_2.
\end{align}

We choose to spread the mass of the droplet $M$ on both masses, $m_{1,2}=M/2$. The length of the spring $L$ is set to 890 $\mu$m which is the typical size of our droplet. The spring constant $k$ models the surface tension $\sigma \approx 0.02$~N/m. This constant $k$ determines the undamped frequency of the spring given by $\omega_0/2\pi=\sqrt{2 k/M}$. Using the damping coefficient $c$, we can  estimate the damping ratio $\xi=\frac{c}{2\sqrt{2 k M}}$ which compares the spring damping effect with the restoring effect. This damping ratio for the spring system is comparable to the Ohnesorge number estimated for the droplet  $\Ohd \approx 0.15$. The droplet spring system is thus underdamped and oscillates at a frequency $\omega_d=\sqrt{2 k(1-\xi^2)/M}$.

This model depends on two parameters only, $k$ and $c$, which are the images of the parameters $\sigma_d$ and $\nu_d$, respectively. We can fine tune them in order to reproduce the bouncing droplet trajectories as accurately as possible. This is done by comparing the numerical simulation of the spring system laid on a static plate with the experimental results of a droplet laid on a static bath, which has been presented in Section~\ref{subsec:ExoObs}.

\subsection{Model parameters}

As we observed that the impact speeds of the bounces on the static bath (cf. Fig.~\ref{fig:DropBounce}) are in the range of the relative impact speeds encountered for droplets bouncing on the oscillating plate (cf. Fig.~\ref{fig:Traj890um50Hz}), we adjusted the model parameters, $k$ and $c$, on the static case. The spring constant $k$ mainly governs the maximal deformation of the droplet during the bounce while the damping coefficient $c$ models the system dissipation and thus the height that the drop reaches at successive bounces. In consequence, The parameters values of the model are set in two steps, the spring constant $k$ is set using the deformation of the droplet at impact and then the damping coefficient $c$ is adjusted using the maximum height reached after bouncing. The best fitting parameters are found to be $k=0.072$~N/m and $c = 112~10^{-6}\,\rm{kg/s}$ , which correspond to $\xi=0.33$. We note that these two parameters do not have independent effects but this method was very satisfying to adjust the parameters in order to reproduce the succession of bounces. We present the result of the numerical simulation using 3 solid lines on Fig.~\ref{fig:DropBounce}. The trajectory of the center of mass of the spring system is represented by the central solid line (red color online) while the trajectories of the masses $m_1$ and $m_2$, assimilated to point masses, are represented by the two outer solid lines (blue color online). Comparing simulations with experiments, we note that the trajectories of the center of mass of the modeled and experimental systems are in good agreement. The additional oscillation of the spring system during the bounce can be attributed to the inertia of the masses fixed at the extremity of the spring. This oscillation has, however, no influence on the next bounce.

\subsection{Numerical simulations}

Considering both calibrated parameters $k$ and $c$ for the model, we performed numerical simulations of the spring system bouncing on a $50\,\rm{Hz}$ oscillating plate.  The values of $c$ and $k$ are not expected to vary with different sizes of the droplet. We adjusted the length of the spring $L$ and the mass of the drop $M$ according to the size of the droplet $D=890~\mu$m that we used in the oscillating bath experiments. This gives a damping ratio for the mass-spring-damper system $\xi=0.25$.

We started by dropping the spring system, from a height of 1~mm, on a rigid plate oscillating at a frequency of $50\,\rm{Hz}$ and with an amplitude such that $\G=1$. The real time simulation was set to 11 seconds ($\approx 550$ oscillations of the plate). We then increased $\G$ by steps of 0.1, without break between the simulations.  We repeated the process until reaching $\G=5$. After that, we decreased $\G$ by steps of 0.1 until reaching $\G=0.5$. We then processed the data of the center of mass of the spring system in the same way as we processed the data displayed selectively on Fig.~\ref{fig:Traj890um50Hz}. For each $\G$ value, we measured each bouncing height $\hreb$ and each time interval $\varDelta T_\mathrm{min}$ between minima of the trajectories of the center of mass.

On Fig.~\ref{fig:DiagBifNumZETA50}c, we reported the different intervals of time $\varDelta T_\mathrm{min}$ between two successive bounces as a function of $\G$. For each $\G$ value (increasing and decreasing phase), $N$ measurements of $\varDelta T_\mathrm{min}$ were binned in classes of 0.1~ms of width (step of the simulation) and $0.05$ of height (corresponding to the experimental error). We defined $n_i$ as the number of events in the bin $i$ corresponding to time within the interval $0.1\,i$ and $0.1\,(i+1)$~ms. On Fig.~\ref{fig:DiagBifNumZETA50}c, the measurement recurrences $n_i/N$ in each bin $i$ are then represented in a grey level corresponding to $(1-n_i/N) \times 255$ where a value of 255 is white and a value of 0 is black. A $100\,\%$ black bin means that all of the measurements corresponding to the acceleration $\G$ are located in the considered bin. Taking bins of 0.1 of width and 0.05~mm of height, we proceeded the same way with the measurements of  $\hreb$ given by the numerical simulation. These are plotted on Fig.~\ref{fig:DiagBifNumZETA50}d. We identified the different bouncing modes on both diagrams and indicated them on each figure.

The agreement between the simulations and the experiments is remarkable. The first bifurcation from (1,1) to (2,2) is located at the same $\Gpdb=1.6$ value. For $\G<\Gpdb$, the two different trajectories characterized by the same mode parameters (1,1),  (1,1)$_H$ and (1,1)$_L$, are observed in the simulations. The (1,1) degenerescence observed experimentally is thus reproduced by the model. The (3,2) mode is also observed and it is overlying the mode (2,1). In fact, the (3,2) mode was observed during the simulations when $\G$ increased and the (2,1) mode was observed in the simulations during the decreasing phases. Thus, depending on the initial conditions one or the other mode can be observed. The bifurcation to the mode (4,2) is also observed for nearly the same acceleration $\G$ in the experiment and simulation. The other mode (2,1) observed experimentally at $\G=4.4$ was detected when analyzing the trajectories of the simulations; however, it was not stable for more than several periods, which is why it does not appear clearly on the figure. The mass-spring-damper system model for the bouncing droplet bouncing is thus able to reproduce the key features observed in the experiments. These features would not have been evidenced by using a bouncing ball approach. This demonstrates that the complexity and the variety of modes that are observed rely on the deformation of the droplet and its finite contact time.

\section{Conclusion and perspectives}

To have a global view of the relative importance of the drop and bath deformations in the bouncing droplets problem, we represented the various works on this problem on a 2D map made with the Ohnesorge number of the bath $\Ohb$ and of the droplet $\Ohd$. In each case, the droplets vertical trajectory can be complex depending on the forcing parameters and various vertical bouncing modes can be observed.  We chose to focus on the role of the droplet deformation on the observed bouncing mode. To highlight its influence, we analyzed small droplets, which can deform as a spring, bouncing on a non-deformable liquid interface. For this purpose, we considered 20~cSt Silicone oil droplet of diameter $D=890$~$\mu$m bouncing on highly viscous liquid bath ($\nu_b=1000$~cSt) that is oscillating at a frequency of 50~Hz. These parameters were chosen in consistency with  recent works on the bouncing droplets \cite{terwagne2010langmuir,dorbolo:2008,gilet:2008, eddi2011vibration,eddi2009archimedean, fort2010memory}.


In these conditions, numerous bouncing modes with multi stable and chaotic regions can be observed when varying the acceleration $\G$ of the oscillating liquid bath. The variety of modes observed is richer than when a low viscous (deformable) bath is used; in this later case the increase of $\G$ is limited by the appearance of the Faraday instability \cite{eddi2008ratchet,protiere2006association}. We should note that the low viscous bath is a necessary condition to perform experiments with droplets interacting with the waves they emit when bouncing on the bath. To have further insight on the role of the droplet deformation on the bouncing modes hierarchy, we made use of bifurcation diagrams in which we can have a global view of the mode succession, the multi stable and the chaotic regions.

To reproduce the complexity and the variety of the bouncing modes, we proposed a model based on a mass-spring-damper system. Indeed, an important feature of the bouncing droplets is that the droplet deforms and stores energy in the surface energy before restoring it and taking off, which process leads to a finite contact time. Moreover, as the bath is accelerating while the droplet is bouncing, a highly non-linear interaction occurs between the bath and the drop. The bouncing ball model fails to capture the behaviors that are observed experimentally. We thus proposed a model that takes into account the deformation of the bouncing droplet  and consequently the finite contact time with bath. It consists in two masses linked by a linear spring of stiffness $k$ in parallel with a dashpot of damping coefficient $c$. Both parameters were adjusted during experiments on a static bath before extending the comparison to an oscillating bath. By simulating numerically the spring system on an oscillating rigid plate allowed us to retrieve the bifurcation diagrams that we observed experimentally. The agreement between the experiments and the simulations is good. This model can even catch the multi stable state like the (1,1)$_H$ mode and the (1,1)$_L$ mode degenerescence. It can also capture both modes characterized by the same parameter (2,1) and observed for different acceleration $\G$. This shows that the mass-spring-damper system (Kelvin-Voigt material) is an excellent model for bouncing droplets.

Finally, using a non-deformable bath allows to isolate the effect of the deformation of the droplet and suggests that a linear spring is suitable to describe the vertical bouncing dynamics of the droplet. When the bath is not deforming, we are able to retrieve most of the modes that are encountered on a deformable bath. Recently, Mol\'a{\v{c}}ek and Bush \cite{molavcek2013bouncing}  modeled the whole system consisting of a deformable drop bouncing on a deformable bath by a single logarithmic spring. In this kind of system, it is rather complicated to isolate the respective contribution of the deformation of the drop or of the bath in the bouncing droplet dynamics. Our study contrasts also with the bouncing droplets on a soap film \cite{gilet:2009a, gilet2009soap} in which it was the deformation of the soap film that was leading to the multi periodic motion of the droplet and not the deformation of the droplet.  So, this raises new questions: by allowing the bath to deform, what is its exact role in the bouncing dynamics of the drop? Does the deformable bath acts like a damper, a spring or both? Our model shall thus be extended to low viscous bath (deformable). Indeed, the coupling between the bath deformation and the droplet bouncing mode are at the origin of fascinating behaviors of the bouncing droplet like the walkers \cite{couder2005nature}, the ratchets \cite{eddi2008ratchet}  and the droplet crystals \cite{eddi2009archimedean}.

\acknowledgments D.T. thanks the Belgian American Education Foundation (B.A.E.F.), the Fulbright Program and the Wallonie-Bruxelles International Excellence Grant WBI.World for financial support. S.D. would like to thank F.R.S.-FNRS for financial support. Part of this work has been supported by COST P21 'Physics of
droplets' (ESF) and the Actions de Recherche ConcertŽes (ARC) "Quandrops". Thanks to Frederic Lebeau for help with the drop dispenser and M. Hubert for a careful reading. \nocite{terwagne:2009, terwagne2007lifetime}


\end{document}